\documentclass[aps,preprint]{revtex4}
\usepackage{epsfig}
\usepackage{amsmath}
\usepackage{epsfig}

\begin{document}
\title
{Hermitian Hamiltonians: Matrix versus Schr{\"o}dinger's}    
\author{$~^1$Zafar Ahmed,$~^2$Mohammad Irfan,$~^3$Achint Kumar,$~^4$Ankush Singhal\\
$~^1$Nuclear Physics Division, Bhabha Atomic Research Centre, Mumbai 400 085, India \\
$~^2$Department of Physics, Indian Institute of Science Education and Research, Bhopal,  462066 India \\ $~^2$Department of Physics, Birla Institute of Technology \& Science, Pilani, Goa, 403726, India, \\
$~^4$Department of Physics, UM-DAE-CBS, Mumbai, 400098, India}
\email{1:zahmed@barc.gov.in, 2:mohai@iiserb.ac.in, 3:achint1994@gmail.com, 4: ankush.singhal@cbs.ac.in}
\date{\today}
\begin{abstract}
We draw attention to the fact that a Hermitian matrix is always diagonalizable and has real discrete spectrum whereas the Hermitian
Schr{\"o}dinger Hamiltonian: $H=p^2/2\mu+V(x)$, may not be so. For instance
when $V(x)=x, x^3, -x^2$, $H$ does not have even one real discrete eigenvalue. Textbooks do not highlight this distinction. However, if $H$ has real discrete spectrum, by virtue of the expansion theorem, one can convert the eigenvalue problem $H\psi_n=E_n \psi_n$ into a matrix and get eigenvalues $E_n$ by diagonalizing the matrix. We show, that the thus obtained $E_n$ could be accurate,  provided $H$ is devoid of scattering states. We suggest that this could be a simple and apt way to introduce the method of Linear Combination of Atomic Orbitals (LCAO) for finding the spectra of molecules. In textbooks, usually the method of matrix-diagonalization appears meagerly as a degenerate perturbation theory for more than one dimensions.
\end{abstract}
\maketitle
Matrices [1,2] are beautiful mathematical entities which date back in  18-19th century have a special place in quantum mechanics which started in the beginning of the 20th century. The latter has borrowed and shared the results from the former but the textbooks do not discuss the subtle 
differences between the two Hamiltonians [3-5]. The present article brings out two dissimilarities between the two and utilizes their similarity to make a simple and apt route to discussing the quantum mechanical method of Linear Combination of Atomic Orbitals (LCAO) at an introductory level. In LCAO [3,4], one calculates the spectrum of  molecules by writing their eigenstates as linear combination of atomic states.

Consider the matrix eigenvalue equation [1,2]: (i) $AU_n=\lambda_n U_n$. If $A$ is Hermitian or real symmetric matrix, it always has real eigenvalues and it is always diagonalizable which means that we can always find another square matrix $D$ such that: (ii) $D^{-1}AD=\Lambda$, where $\Lambda$ matrix is a diagonal matrix having  $[\lambda_1,\lambda_2,\lambda_3,...,\lambda_n]$  as diagonal elements and $D$ is composed by collecting $n$ linearly independent eigenvectors $U_n$.  Here $\lambda_n$ are real eigenvalues which are the roots of the determinantal equation: (iii) $\det |A-\lambda I|=0$ .  Additionally the eigenvectors are orthonormal:  (iv) $U_n^\dagger U_k=\delta_{n,k}$ and complete: (v) $\sum_{k=1}^n U_k U^\dagger_k={\cal I}_{N \times N}$. All these properties are irrespective of whether eigenvalues/roots $\lambda_n$ are repeated or not.

Other types of square (non-Hermitian) matrices (see Appendix)  may also have real eigenvalues and they are diagonalizable if they have distinct eigenvalues or in case of $m$ repeated roots, if one can construct $m$ linearly independent non-null vectors. More interestingly their eigenvectors will not be orthonormal and complete as discussed above (see (iv) and (v) above ).  

\begin{figure}[ht]
\centering
\includegraphics[width=5 cm,height=5 cm]{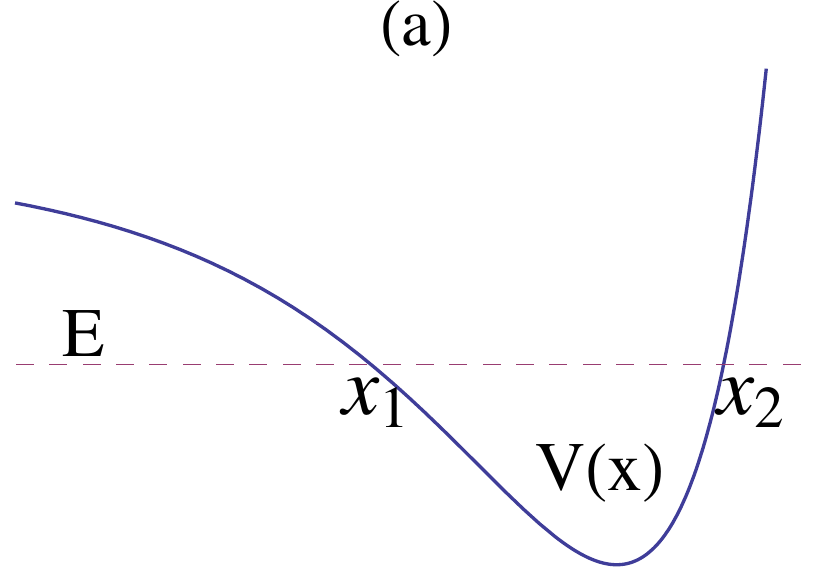}
\hskip .5 cm
\includegraphics[width=5 cm,height=5 cm]{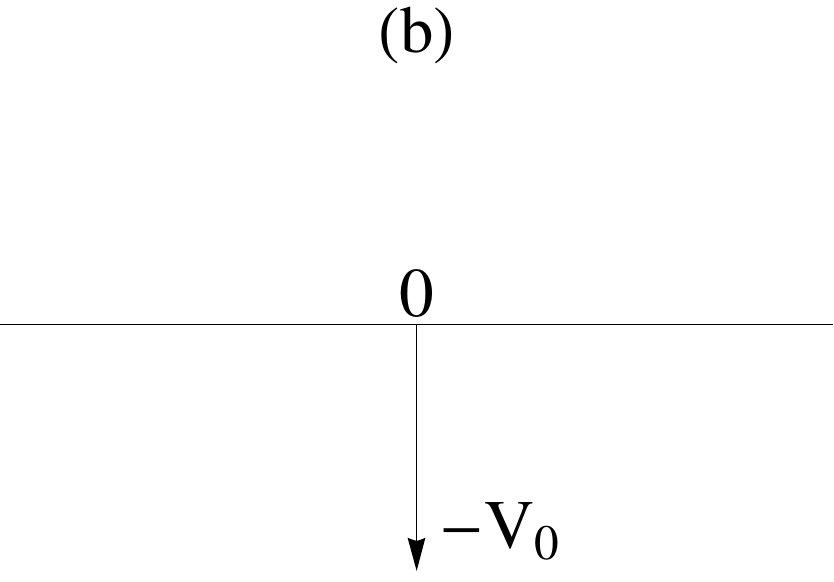}
\hskip .5 cm
\includegraphics[width=5 cm,height=5 cm]{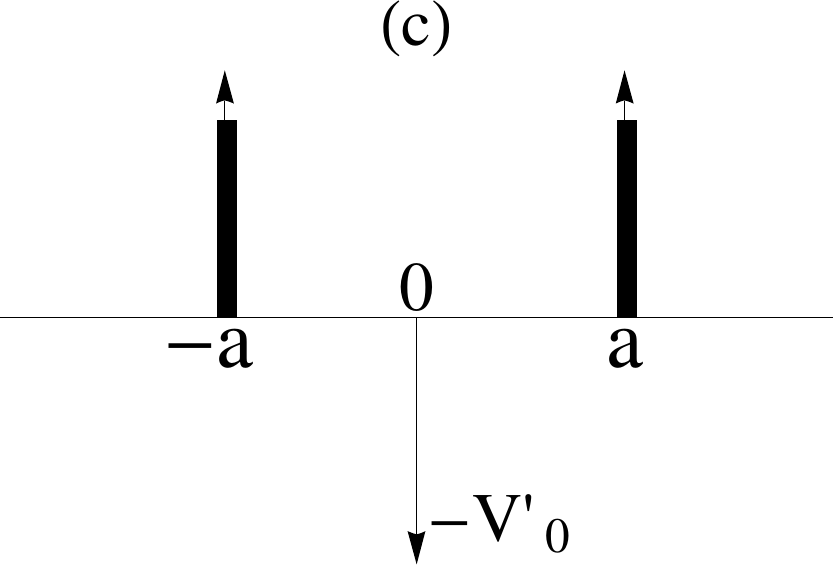}
\caption{Schematic depiction of various potentials (a) : a potential well $V(x)$ with real pairs of turning points $x_1$ and $x_2$ at an energy $E$, (b): The Dirac Delta well $V(x)=-V_0 \delta(x)$, (c): V'(x), which is the Dirac Delta well between two rigid walls}
\end{figure}

In quantum mechanics eigenvalue equation: $H\psi_n=E_n\psi_n$ was introduced by Schr{\"o}dinger in the year 1925-26. Here $H=p^2/2\mu +V(x)$ and $p=-i\hbar
\frac{d}{dx}$ is a differential operator which converts the eigenvalue equation into a second order differential equation
\begin{equation}
\frac{d^2\psi_n(x)}{dx^2}+\frac{2\mu}{\hbar^2}[E_n-V(x)]\psi_n(x)=0,
\end{equation}
where for discrete eigenvalues one has to impose the Dirichlet Boundary Condition (DBC) that $\psi_n(\pm \infty)=0$. Since allowed energy values have to be real so here we demand $V(x)$ to be real  (Hermitian in a simple way). In text books it is generally proved that real  potential along with DBC will have real eigenvalues. However, the fact that  only specific profiles of $V(x)$ can  possess eigenvalues is ignored.

It is here we emphasize the use of a series of semi-classical ideas that can  decide {\it a priori} whether $V(x)$ will have any discrete eigenvalue. These semi-classical pointers are: $\{1\}~ V(x)$ needs to have a minimum. $\{2\}$ For $\tilde E$ to be an eigenvalue, the equation $\tilde E=V(x)$ needs to have two real roots $x_1, x_2$ called classical turning points. \{3 \} Next, $\tilde E>V(x)$  $\forall ~ x \in (x_1,x_2)$. $\{4\}$ (see Fig. 1(a)). Finally, the action integral $I$
\begin{equation}
\hbar^{-1}\int_{x_1}^{x_2} \sqrt{2\mu[\tilde E- V(x)]} dx
\end{equation}
has to take one of the values $(n+1/2), n=0,1,2,3..$. These conditions also are only necessary but not sufficient. Despite meeting these many conditions the Schr{\"o}dinger equation  may allow $E=\tilde E \pm \epsilon$ as an eigenvalue of $V(x)$, making $\tilde E$ to be only an approximate eigenvalue. It may be readily checked that $V(x)=x^{2m+1},-x^{2m} (m =1,2,...)$ cannot have discrete energy  bound states but $V(x)=x^{2m}$ do have them

Thus, for real discrete spectrum unlike matrices for Schr{\"o}dinger Hamiltonians, the condition of Hermiticity [6] may not be sufficient. Another difference between  the two is that a matrix  possesses only discrete spectrum and it does not have continuous spectrum. However, the Schr{\"o}dinger's Hamiltonian can have both types of spectra:  discrete and continuous.

Red, Green and Blue (RGB) are supposed to be fundamental (basic) states of colors so any linear combination of them say  $C=xR+yG+zB$ creates a new color if $x,y,z\ge 0$ and $x+y+z=1$. So any color is a vector $(x,y,z)$ in the space of RGB. Since every color could be codified as a vector (information) so it could be transmitted and consequently  we have color TVs. On the other hand, a taste or a smell cannot be transmitted as we do not know their basic states. Every bit of music, we create or hear, it seems it is some  combination of the fundamental notes  of octave.  In mathematics, any real and finite function $f(x)$ defined in $[-L,L]$ could be represented very well excepting at the end points as a linear combination of $\sin n \pi \frac{x}{L}$ and $\cos n  \pi \frac{x}{L}$. This linear combination is known as Fourier series [1]:
\begin{equation}
f(x)=\frac{A_0}{2} +\sum_{n=1}^{\infty}[ A_n \cos n \pi \frac{x}{L} +B_n \sin n \pi \frac{x}{L}], \quad n=1,2,3...
\end{equation}
so $f(x)$ will be well represented by $\{A_0,A_n, B_n\}$, in $x \in (-L,L)$, the basic states
$\sin n \pi \frac{x}{L}, \cos n \pi \frac{x}{L}$ are orthogonal as:
\begin{equation}
\int_{-L}^{L} \sin m \pi \frac{x}{L}~  \sin n  \pi \frac{x}{L} ~dx= L \delta_{m,n} = \int_{-L}^{L} \cos m  \pi \frac{x}{L} ~ \cos n \pi \frac{x}{L} ~ dx,  \int_{-L}^{L} \sin m \pi \frac{x}{L}~ \cos n \pi \frac{x}{L} ~dx
= 0.
\end{equation}
$A_n$ and $B_n$ are calculated as
\begin{equation}
A_n=\frac{1}{L} \int_{-L}^{L} f(x) \cos n\pi \frac{x}{L}~ dx, \quad
B_n=\frac{1}{L} \int_{-L}^{L} f(x) \sin n\pi \frac{x}{L}~ dx.
\end{equation}
In quantum mechanics suppose a particle is in a potential well $V(x)$. Let its bound state eigenvalues and eigenfunctions  be  $\epsilon_n$ and  $\phi_n(x)$, respectively;  such that $H \phi_n(x)= \epsilon_n \phi_n(x)$.
Now any time the state $\Psi(x,t)$ which is the solution of the time-dependent Schr{\"o}dinger equation 
\begin{equation}
[H -i\hbar \frac{\partial}{\partial t}]\Psi(x,t)=0,
\end{equation}
can be written as [3-5]
\begin{equation}
\Psi(x,t)= \sum_{n=0}^{\infty} c_n ~e^{-i\epsilon_n t/\hbar}~\phi_n(x),
\end{equation}
provided $H$ is time independent. When time dependence in not required, any state $\psi(x)$ of the particle in $V(x)$ can be written as [3-5]
\begin{equation}
\psi(x)= \sum_{n=0}^{\infty} c_n \phi_n(x).
\end{equation}
The Eqs. (7) and (8) are known as the expansion theorem of quantum mechanics.
Somehow these two statements  do not actually glorify the expansion theorem as much as it should be. On par with the Fourier theorem, we can say that any
eigenstate $\psi_n(x)$ of a Hamiltonian ${\cal H}$ can be expanded as a linear combination of the complete orthogonal eigenstates $\phi_n(x)$ of the Hamiltonian $H$ where $H \phi_n(x)= \epsilon_n \phi_n$. So in the eigenvalue equation 
\begin{equation}
{\cal H}\psi(x)= E \psi(x), 
\end{equation}
let us put $\psi(x)=
\sum_{k=0}^N C_k \phi_k(x)$ to write
\begin{equation}
\sum_{k=1}^{N} C_k{\cal H}  \phi_k(x) = E \sum_{k=1}^N C_k \phi_k(x).
\end{equation}
Multiplying both sides by $\phi_n(x)$ and integrating w.r.t $x$, from $-\infty$  to $\infty$, we get
\begin{equation}
\sum_{k=1}^{N}C_k \int_{-\infty}^{\infty} \phi_n (x) {\cal H} \phi_k(x)~dx=
E \sum_{k=1}^{N} C_k \int_{-\infty}^{\infty} \phi_n (x)  \phi_k(x)~dx. 
\end{equation}
using the orthonormality condition that $\int_{-\infty}^{\infty} \phi_n(x) \phi_k(x)~dx= \delta_{n,k}$, we can write
\begin{equation}
\sum_{k=1}^{N} [{\cal H}_{n,k} - E \delta_{n,k}] C_k=0,
\end{equation} 
where ${\cal H}_{n,k}=\int_{-\infty}^{\infty} \phi_n(x) {\cal H}  \phi_k(x) ~dx$.
Now if we run $n$ from 1 to $N$, these are $N$ linear simultaneous 
homogeneous equations  in $N$ variables $C_1,C_2,C_3,...C_N$ :  ${\cal M} {\cal C}={\cal O}$, where ${\cal M}$ is $N \times N$ matrix and ${\cal C}$ is a column vector. This equation will have non-zero and non-trivial solutions when $\det|{\cal M|}=0$, this gives the secular or characteristic equation 
\begin{equation}
\det|{\cal H}-E {\cal I}|_{N \times N}=0. 
\end{equation}
According to the Fundamental Theorem of Algebra a polynomial with real coefficients
has either real or non-real complex conjugate roots.
This determinantal equation (13) despite being  a real polynomial equation in $E$ 
will have all $N$  roots as real by virtue of the fact that ${\cal H}$ is Hermitian matrix. These roots are nothing but approximate eigenvalues $E_n$ of (9) and these will become more and more accurate when we increase $N$. Since in this method we are considering the complete discrete energy eigenstates $\phi_n(x)$ of the Hamiltonian $H$ assuming that the Hamiltonian ${\cal H}$ does not have continuous energy (scattering states) spectrum  so this method is expected to be  accurate for the Hamiltonians ${\cal H}$ not having scattering states. In other cases this method would be approximate.

In textbooks, usually an equation identical to (13) appears in a restricted sense namely as the degenerate perturbation theory [3-5] for more than one dimensions. For instance, the first excited state ($n=2$)  of hydrogen atom is four fold-degenerate; the quantum numbers $l,m$ have the values (0,0), (1,0), (1,1) and (1,-1). In stark effect, when the hydrogen atom is perturbed by an external electric field $H^\prime= e {\cal E} r \cos \theta$; a linear combination of the electronic states of hydrogen  atom is taken as $\chi=C_1 ~\chi_{0,0} + C_2 ~ \chi_{0,1}+ C_3~ \chi_{1,1} + C_4~ \chi_{1,-1}$ as a  solution of the total Hamiltonian $(H_{hy}+H^\prime)$. Then the diagonalization of the $4 \times 4$ yields four eigenvalues $0,0, +3e{\cal E}a_0,  -3e{\cal E}a_0$, consequently half of the four-fold degeneracy is removed [5]. In distinction to this, in this paper, we are professing the  calculation of eigenvalues of a one-dimensional Hamiltonian by the matrix- diagonalizaton. Though, at research levels, this is  practiced commonly but in textbooks it is presented meagerly only as a degenerate perturbation theory.

Fortunately, in one-dimensional quantum mechanics, we have two complementary systems  having complete set of bound eigenstates, these are a particle between two rigid walls at $x=\pm a$ (finite support, Infinitely Deep Well (IDW)) and the other one is of infinite support called Harmonic oscillator (HO). In following discussions and calculations we work with units where we set $2\mu=1=\hbar^2$.  For the IDW, the normalized basic states are [3-5]
\begin{equation}
\phi_n(x)=|n>= \sqrt{\frac{1}{a}}\cos \frac{n \pi x}{2a} \quad  (n-\mbox{ odd}), \quad \sqrt{\frac{1}{a}} \sin \frac{n \pi x}{2a}  \quad (n-\mbox{ even}), \quad E_n=\frac{n^2 \pi^2}{4a^2}.
\end{equation}
For HO: $H=p^2+x^2$ ($\omega=2$), the normalized basic states are [3-5]
\begin{equation}
\Phi_n(x)=|n>=(2^n n! \sqrt{\pi})^{-1/2}~ e^{-x^2/2}~ H_n(x), \quad E_n=2n+1.
\end{equation}
In this paper, we consider two Hamiltonians ${\cal H}=p^2+V(x)$ and ${\cal H}^\prime= p^2+V^\prime(x)$, where $V(x)=-V_0 \delta(x)$  (Fig. 1(b)) and $V^\prime(x)$ is the same Dirac delta potential well between two rigid walls  (Fig. 1(c)). It is known that ${\cal H}$ has  one discrete eigenvalue ($E_0=-V_0^2/4$) bound state along with  positive energy continuum of scattering states [3]. But ${\cal H}^\prime$ has only discrete energy spectrum of infinitely many bound states [7]. The energy  eigenvalues of $V'(x)$ are given [6]  by $\tan ka=2k/V'_0$ (even parity states), $ka=n\pi$ (odd parity states). When $V'_0 a=2$, $E=0$ is an eigenstate. When $V'_0  a>2$ the negative energy bound state is found by solving $\tanh \kappa a=2\kappa/V'_0$. Here $k=\sqrt{E}$ and $\kappa=\sqrt{-E}.$

In the IDW basis we have
\begin{eqnarray}
<m|p^2|n>=\int_{-a}^{a} \phi_m(x) p^2 \phi_n(x) dx=\frac{m^2 \pi^2}{4a} \delta_{m,n}, \quad
<m|-V_0\delta(x)|n>= \\ \nonumber -V_0 \int_{-a}^{a} \phi_m(x) \delta(x) \phi_n(x) ~ dx  =-\frac{V_0}{a} ~ \eta_{m,n},
\end{eqnarray}
where $\eta_{m,n}=\phi_m(0)\phi_n(0)=1$, if both $m$ and $n$ are odd integers, otherwise it is 0.
So the matrix element ${\cal H}_{m,n}$ for the Hamiltonian of Dirac delta potential between two rigid walls (Fig. 1(c)) in IDW basis can be written as
\begin{equation}
{\cal H}^\prime_{m,n}=\frac{m^2\pi^2}{4a} \delta_{m,n}-\frac{V_0}{a}~\eta_{m,n}.
\end{equation}
For HO basis by using the creation and annihilation operators [3-5], we can write
\begin{equation}
<m|p^2|n>=\frac{1}{2}[(2n+1)~\delta_{m,n}-\sqrt{n(n-1)}~\delta_{m,n-2}-\sqrt{(n+1)(n+2)}~\delta_{m,n+2}],
\end{equation}
and
\begin{equation}
<m|-V_0 \delta(x)|n>=-V_0 ~\int_{-\infty}^{\infty} \Phi_m(x) \delta(x)  \Phi_n(x)~dx=-V_0 ~\Phi_m(0)~ \Phi_n(0)= -V_0 \xi_{m,n},
\end{equation}
where utilizing the available [8] nonvanishing values of even ordered Hermite polynomials $H_{2k}(0)$, we obtain
\begin{eqnarray}
\xi_{m,n}&=&\frac{2^{(m+n)/2}\sqrt{\pi}}{\sqrt{m!n!}~ \Gamma[(1-m)/2]~\Gamma[(1-n)/2]}, \quad \mbox{if both}~ m ~\mbox{and}~ n ~ \mbox{are even} \\ \nonumber
&=&0, \quad\mbox{otherwise}.
\end{eqnarray}
So the matrix elements for the Dirac delta  well (Fig. 1(b)) can be written as
\begin{equation}
{\cal H}_{m,n}=<m|p^2|n>-V_0 \xi_{m,n}.
\end{equation}
In order to find the discrete spectrum of ${\cal H}^\prime$ we diagonalize the matrix
${\cal H}^\prime_{N \times N}$ by taking $N$ as 10, 50, 100, 500 and 1000 and follow first four eigenvalues for four values of $V_0^\prime$. For all the cases we fix $a=1$. See in the Table that $E_0$ and $E_2$ show fast convergence  as $N$ increases from 10 to 1000.  For $N=1000$ or much less values these eigenvalues match with the exactly calculated eigenvalues accurately upto two decimal places. This means that for a better accuracy $N$ needs to be increased further. Interestingly, for all the values of $V_0^\prime$, $E_1$ and $E_3$ coincide with exact values $\pi^2$ and $4\pi^2$ which are independent of $V_0^\prime$ as the eigenvalues come from the condition $ka=n\pi$.

Contrary to this, for ${\cal H}$ the diagonalization of ${\cal H}_{N \times N}$ displays slow convergence of its only eigenvalue $(E_0)$ as $N$ is increased from 10 to 1000 up to even two decimal places  and increasing $N$ further does not change $E_0$ significantly. Eventually, the exact value of $E_0=-V_0^2/4$ is not obtainable.
This we attribute to the presence of scattering states over the attractive Dirac Delta potential for the continuum of positive energies. This is the characteristic feature of the matrix  as they are insensitive to the scattering states.  It is instructive to remark that exact value of $E_0=-(\lambda-g)^2/4$ has been obtained when a  delta potential $-\lambda \delta(x)$ was perturbed by a delta  barrier $g\delta(x)$ using  the contribution of scattering states [9]
in the second order perturbation theory. 

The diagonalization of a $N \times N$ matrix gives $N$ eigenvalues even if the potential has just one or  a finite number of discrete eigenvalues. In these cases 
we have to consider the lowest eigenvalues as the correct ones. Interestingly, for ${\cal H}$, only the first eigenvalue turns out to be negative for any value of $N$.
Even if  a Schr{\"o}dinger Hamiltonian does not possess discrete spectrum but diagonalization of its corresponding matrix will yield $N$ eigenvalues. Now the question as to how to discard these (so called) eigenvalues is intriguing. We find that either every eigenvalue does not show the convergence as we increase $N$ or all the eigenvalues converge to zero when $N$ is made very large.

Yet another point to ponder is that in a Hermitian matrix the eigenvalues may repeat
(degeneracy) or not. However, for  the Hermitian Schr{\"o}dinger Hamiltonian the degeneracy is essentially absent in one-dimension, it may occur in two or three dimensions.  This is why, whenever we obtain the eigenvalues of a one dimensional Hamiltonian by the method of matrix diagonalization  eigenvalues do not repeat. For higher dimensions, in textbooks the method of diagonalization is presented meagerly only as a degenerate perturbation theory. In this regard, the example of Stark effect [3-5] of hydrogen atom is outlined above where $E=0$ repeats twice.

To summarize, in this paper we have shown that for real discrete eigenvalues  the condition of Hermiticity is sufficient only for a matrix Hamiltonian but it may not be so for a Schr{\"o}dinger Hamiltonian. We have pointed out that matrices unlike 
Schr{\"o}dinger Hamiltonians do not possess continuous spectra. Based on the expansion theorem we have shown that if a  Schr{\"o}dinger Hamiltonian possess discrete spectrum of bound states, it can be converted to a matrix of order $N \times N$  whose diagonalization for a sufficiently large dimension ($N$) can yield the discrete spectrum  accurately, provided the Schr{\"o}dinger Hamiltonian does not entail scattering states. Otherwise this method would be approximate. We find that this is the primary route to introduce the method of Linear Combination of Atomic Orbital (LCAO).

\section*{Appendix}
The non-Hermitian  matrix $A=\left (\begin{array}{clcr} 1 &1\\ 4 & 1 \end{array} \right)$ is diagonalizable. But the non-Hermitian $B=\left (\begin{array}{clcr} 1 &1\\ 0 & 1 \end{array} \right)$ [3] is not diagonalizable. Following are the examples  of $3 \times 3$ non-Hermitian matrices which have real discrete spectra and they may or may not be diagonalizable. \\

\underline{Non-Hermitian matrix with distinct eigenvalues: Diagonalizable}\\ \\
$C=\left (\begin{array}{clcr} 
1 & 0 & 0\\ 0 & 2 & 1 \\ 2 & 0 & 3 
\end{array}\right),\quad U_1=\left(\begin{array}{clcr} 0 \\1 \\1 \end{array}\right), \quad U_2=\left(\begin{array}{clcr} 0 \\1 \\0 \end{array}\right), \quad U_3=\left(\begin{array}{clcr} -1 \\-1 \\1 \end{array}\right),\quad  \lambda_k= 3,2,1.$ \\

\underline{Non-Hermitian matrix with 2 repeated eigenvalues: Diagonalzable (case)}\\ \\
$D=\left (\begin{array}{clcr} 
3 & 1 & 1\\ 2 & 4 & 2 \\ 1 & 1 & 3 
\end{array}\right),\quad U_1=\left(\begin{array}{clcr} 1 \\2 \\1 \end{array}\right), \quad U_2=\left(\begin{array}{clcr} -1 \\0 \\1 \end{array}\right), \quad U_3=\left(\begin{array}{clcr} -1 \\ 1 \\0 \end{array}\right),\quad  \lambda_k= 6,2,2.$ \\ 

\underline{Non-Hermitian matrix with 2 repeated eigenvalues: Non-diagonalizable(case)}\\ \\
$E=\left (\begin{array}{clcr} 
3 & 10 & 5\\ -2 & -3 & -4 \\ 3 & 5 & 7 
\end{array}\right),\quad U_1=\left(\begin{array}{clcr} -1 \\-1 \\2 \end{array}\right), \quad U_2=\left(\begin{array}{clcr} 5 \\2 \\-5 \end{array}\right), \quad U_3=\left(\begin{array}{clcr} 0 \\ 0 \\0 \end{array}\right),\quad  \lambda_k= 3,2,2.$ \\

Notice that the third eigenvector would be proportional to $U_2$ or it could also be a null vector, in any case the diagonaizing matrix ${\cal D}$ made by such three vectors is singular and ${\cal D}^{-1}$ does not exist.
\begin{table}[t]
\caption{First four eigenvalues of $V^\prime(x)$ and $V(x)$ potentials  for various values of $V^\prime_0, V_0$ and N.}
\begin{tabular}{|c|c|c|c|c|c||c|c|c|}
\hline
~~~$V^\prime_0$~~~~~~ &~~~ N~~~~~~  &  $E_0$  & $E_1$ & $E_2$ & $E_3$ & $V_0$  & N & $E_0$ \\
\hline
~~4~~ &~~10~~&~~ -3.0294~~ &~~9.8696~~& ~~18.5529~~ & ~~39.4784 & 4 & 10& -2.5642 \\
~&~~50~~ &~~ -3.5277~~ &~~9.8696~~& ~~ 18.3315 ~~ & ~~39.4784~~&  ~ & 50 & -3.1705 \\
~&~~100~~&~~-3.5967~~& ~~9.8696~~& ~~18.3028~~& ~~39.4784~~& ~& 100 & -3.3760 \\
~&~~500~~& -3.6530 & 9.8696 & 18.2796& 39.4784 & ~& 500 & -3.6974 \\
~&~~1000~~ & -3.6601& 9.8696 &  18.2767 & 39.4784 &~& 1000 &-3.7819 \\
~& Exact & -3.6672 & $\pi^2$ & 18.2738 & $4\pi^2$ & ~ & Exact & -4 \\	
\hline
~~3~~& 10 & -1.3512 & 9.8696 & 19.3773 & 39.4784 &  3 & 10 & -1.5919 \\
~& 50 & -1.5929 & 9.8696 &  19.2422 & 39.4784 & ~& 50& -1.8842 \\
~&100 & -1.6255 & 9.8696 & 19.2249 & 39.4784 & ~ & 100 & -1.9781 \\
~& 500 & -1.6518 & 9.8696 & 19.2109 & 39.4784 & ~& 500 &-2.1204 \\
~& 1000 & -1.6552 & 9.8696 & 19.2092 & 39.4784 & ~& 1000& -2.1570 \\
~& Exact & -1.6585 & $\pi^2$ & 19.2074 & $4\pi^2$ & ~& Exact & -9/4 \\
\hline
~2~ & 10 & 0.1155& 9.8696& 20.2706& 39.4784 & 2 &10&-0.7853 \\
~& 50 & 0.0240 & 9.8696 & 20.2068 & 39.4784 & ~& 50 & -0.8866 \\
~&100& 0.0120& 9.8696& 20.1988& 39.4784& ~& 100 &-0.9167 \\
~& 500 & 0.0024 & 9.8696 & 20.1923 & 39.4784 & ~& 500 & -0.9610 \\
~& 1000 & 0.0012 & 9.8696 & 20.1915 & 39.4784 & ~& 1000&-0.9610 \\
~& Exact & 0 & $\pi^2$ & 20.1907 & $4\pi^2$ & ~& Exact&-1 \\
\hline
~1& 10 & 1.3827 & 9.8696 & 21.2194 & 39.4784& 1 & 10&-0.2097 \\
~& 50 & 1.3634 & 9.8696 & 21.2029 & 39.4784&  ~& 50 & -0.2350 \\
~&100& 1.3610& 9.8696 & 21.2009& 39.4784& ~& 100 & -0.2392 \\
~& 500 & 1.3590 & 9.8696 & 21.1992& 39.4784&  ~& 500 & -0.2450 \\
~& 1000 & 1.3587 & 9.8696 & 21.1990 &  39.4784&  ~& 1000 &-0.2464 \\
~& Exact & 1.3585 & $\pi^2$ & 21.1988& $4\pi^2$ & ~& Exact& -1/4 \\
\hline
0&10 & 2.4674& 9.8696& 22.2066 & 39.4784& -& -& -\\
~& 1000 & 2.4674 & 9.8696 & 22.2066 & 39.4784& -&-&-\\
~ &Exact & $\pi^2/4$ & $\pi^2$ & $9\pi^2/4$ & $4\pi^2$ & -& -&-\\
\hline
\end{tabular}
\end{table}

\end{document}